# FUZZY LOGIC BASED DIRECT TORQUE CONTROL OF INDUCTION MOTOR WITH SPACE VECTOR MODULATION

Fatih Korkmaz[1], İsmail Topaloğlu[2] and Hayati Mamur[3]

[1,2,3] Department of Electrical and Electronics Engineering
Çankırı Karatekin University, 18200, Çankırı, TURKEY

***ABSTRACT***

*The induction motors have wide range of applications for due to its well-known advantages like brushless structures, low costs and robust performances. Over the past years, many kind of control methods are proposed for the induction motors and direct torque control has gained huge importance inside of them due to fast dynamic torque responses and simple control structures. However, the direct torque control method has still some handicaps against the other control methods and most of the important of these handicaps is high torque ripple. This paper suggests a new approach, Fuzzy logic based space vector modulation, on the direct torque controlled induction motors and aim of the approach is to overcome high torque ripple disadvantages of conventional direct torque control. In order to test and compare the proposed direct torque control method with conventional direct torque control method simulations, in Matlab/Simulink, have been carried out in different working conditions. The simulation results showed that a significant improvement in the dynamic torque and speed responses when compared to the conventional direct torque control method.*

## 1. INTRODUCTION

In many industrial applications, Direct torque control (DTC) of induction motor is well-known control method which provides fast dynamic response compared with other control methods like field oriented control (FOC). The DTC has been proposed for induction motor control in 1985 by Takahashi [1] and similar idea that the name of Direct Self Control devoloped in 1988 by Depenbrock [2].

Over the past years, the DTC has gained great attention due to its advantages like simple control structure, robustness to parameters variations, fast dynamic response, not need to current regulators...etc. However, DTC has still some disadvantages and they can be summarized as follows; high current and torque ripples, difficulty to control torque and flux at very low speed, variable switching frequency behavior and high sampling frequency needed for digital implementation.

On the other hand, intelligent control methods like fuzzy logic have been explored by several researchers for its potential to incorporate human intuition in the design process. The FL has gained great attention in the every area of electromechanical devices control due to no need mathematical models of systems unlike conventional controllers[3]. A fuzzy logic controller is used to select voltage vectors in conventional DTC in some applications [4-6]. In parameter estimation applications, a fuzzy logic stator resistance estimator is used and it can estimate changes in stator resistance due to temperature change during operation[7]. For duty ratio control method, a fuzzy logic controller is used to determine the duration of output voltage vector at each sampling period [8]. These fuzzy logic controllers can provide good dynamic performance and robustness. In recent publications, we see that some flux optimization method is proposed for the DTC method for induction motor drives and the effects of the optimization algorithm is investigated. In these publications, three flux control methods are used for optimization and essentially we can classified according to control structure as following: flux control as a function of torque [9], flux control based on loss model [10] and flux control by a minimum loss search controller [11].

In this paper, to reduce the torque ripples of the induction motor on the DTC method, a new approach has been proposed which named as, fuzzy logic based space vector modulation method(FL-SVM). The fuzzy logic controller, in this proposed method, rates of flux and torque errors as input and describes optimum space vector as output to minimize flux and torque errors. The numerical simulation studies have performed with Matlab/Simulink to performance testing of the proposed control method.

## 2. DIRECT TORQUE CONTROL

The basic idea of the DTC technique is to choosing the optimum vector of the voltage, which makes the flux rotate and produce the desired torque. In conventional DTC method, the control of an induction motor involves the direct control of stator flux vector by applying optimum voltage switching vectors of the inverter. For this control, the stator current should be decoupled two independent components as flux and torque components like dc motors. The clarke transformation method is uses in this decoupling process in the DTC method. The DTC allows for very fast torque responses, and flexible control of the induction motor.

The DTC bases on the selection of the optimum voltage vector which makes the flux vector rotate and produce the demanded torque. In this rotation, the amplitude of the flux and the torque errors are kept within acceptable limits by hysteresis controllers [12]. The rotation of the stator flux vector and an example for the effects of the applied inverter switching vectors are given in the Fig. 1.

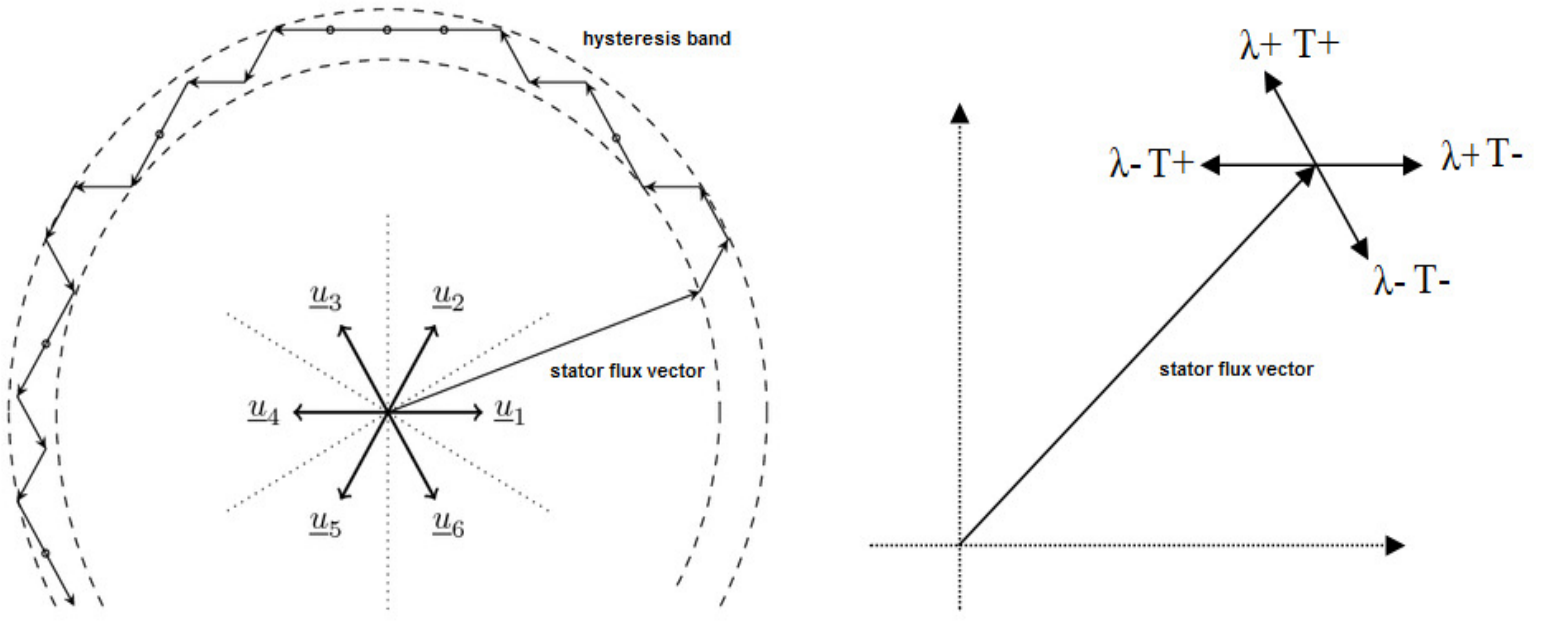

Figure 1. The rotation of the stator flux vector and an example for the effects of the applied inverter switching vectors

The stator flux linkage vector is estimated by using Eq. (1)-(3).

$$\lambda_\alpha = \int (V_\alpha - R_s i_\alpha) dt, \tag{1}$$

$$\lambda_\beta = \int (V_\beta - R_s i_\beta) dt, \tag{2}$$

$$\lambda = \sqrt{\lambda_\alpha^2 + \lambda_\beta^2}. \tag{3}$$

Where $\lambda$ is stator flux vector, $v_\alpha$ and $v_\beta$ stator voltages two phase components, $i_\alpha$ and $i_\beta$ line currents in α-β reference frame and $R_s$ describes stator resistance. After the calculation of the α-β components of the stator flux vector electromagnetic torque of the induction motor can be calculated as given in Eq. 4.

$$T_e = \frac{3}{2} p(\lambda_\alpha i_\beta - \lambda_\beta i_\alpha). \tag{4}$$

Where, $p$ is the number of pole pairs. In DTC method, stator flux rotate trajectory devided six region and well-defined of stator flux region is directly affects on control performance. The stator flux α-β components that calculated can be use the defining of the stator flux region as given in Eq. 5.

$$\theta_\lambda = \tan^{-1}(\frac{\lambda_\beta}{\lambda_\alpha}). \tag{5}$$

These observed values of the flux and the torque errors are compared to reference the flux and the torque values and the resultant errors are applied to the hysteresis comparators as inputs. Two different hysteresis comparators, as flux and torque comparators, generate other control parameters on the DTC method. According to the hysteresis comparators outputs, the observed angle of flux linkage and using a switching table, optimum voltage vectors are selected and applied to the inverter.

## 3. BASICS OF THE FUZZY LOGIC BASED DTC

The objective of space vector pulse width modulation technique is to obtain the demanded output voltage by instantaneously combination of the switching states corresponding the basic space vectors (Figure 2.)[13].

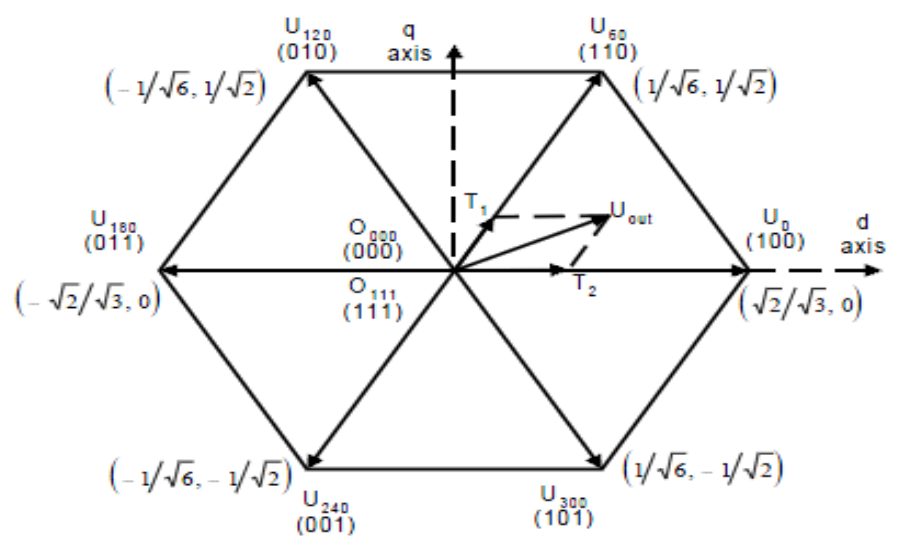

Figure 2. Basic space vectors

$U_{out}$ can be obtained as Eq. 6. by applying the inverter in switching states $U_x$ and $U_{x+60}$ or $U_{x-60}$ for the time periods, $T_1$ and $T_2$ periods of time respectively.

$$U_{out}(nT) = \frac{1}{T}(T_1 U_x + T_2 U_{x\pm 60}) \tag{6}$$

It must be pointed out that the sum of $T_1$ and $T_2$ periods should be less than or equal to total sampling time period, T. If $T_1 + T_2 \langle T$, than the inverter needs to be in pasive vectors, $0_{000}$ or $0_{111}$ states, for the rest of the total time period that pasive time period can be named $T_0$. Thus, the calculation of total time period is given in Eq. 7.

$$T_1 + T_2 + T_0 = T \tag{7}$$

Therefore, Eq. 6 becomes Eq. 7 in the following,

$$TU_{out} = T_1 U_x + T_2 U_{x\pm 60} + T_0 (0_{000} or 0_{111}) \tag{8}$$

From Eq. 8., we get Eq. 9. for $T_1$ and $T_2$.

$$[T_1 \quad T_2]^{\tau} = T[U_x \quad U_{x\pm 60}]^{-1} U_{out} \tag{9}$$

where $[U_x \quad U_{x\pm 60}]^{-1}$ is the normalized decomposition matrix for the sector.

Assume the angle between $U_{out}$ and $U_x$ is $\alpha$ from Figure 3, it can be also obtained Eq. 10. and Eq. 11. for the $T_1$ and $T_2$ [13].

$$T_1 = \sqrt{2}T \| U_{out} \| \cos(\alpha + 30°) \tag{10}$$

$$T_2 = \sqrt{2}T \| U_{out} \| \sin(\alpha) \tag{11}$$

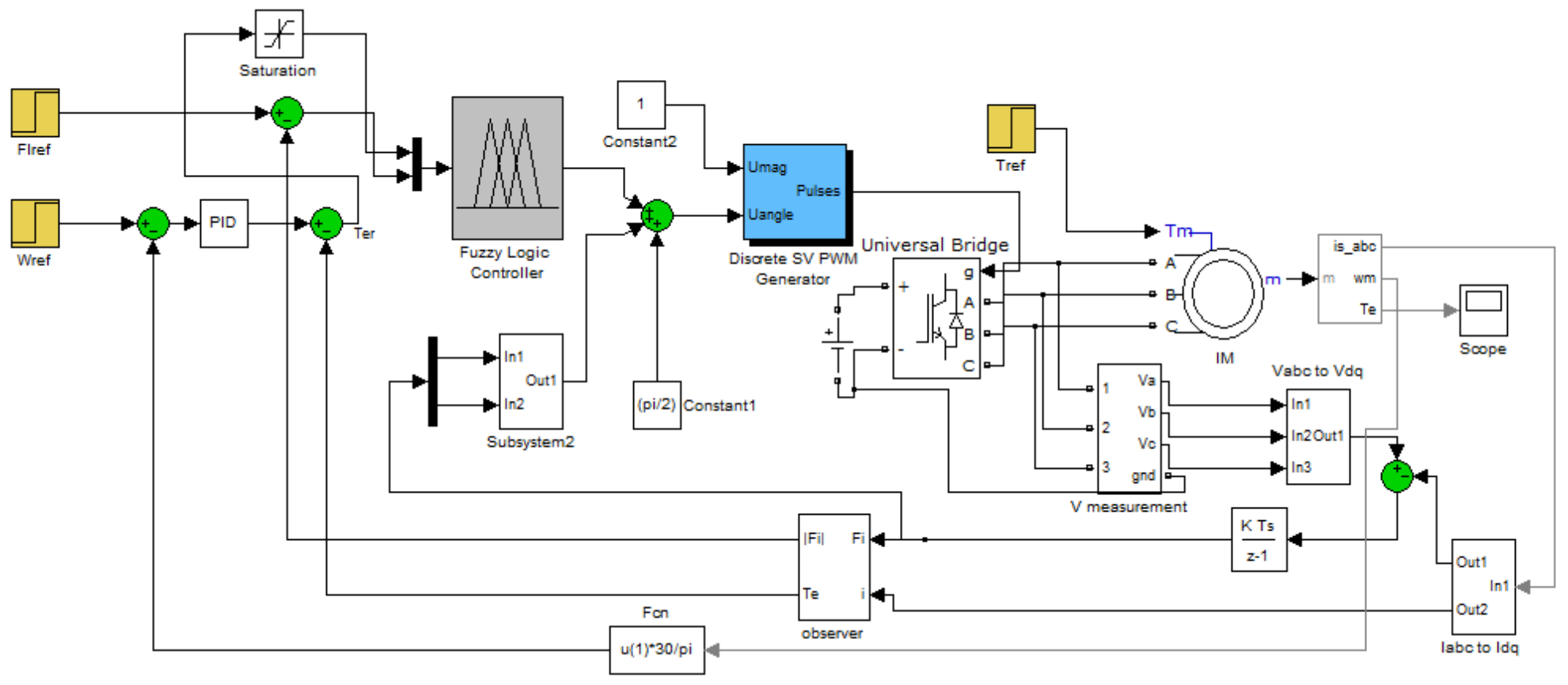

Figure 3. The Simulink block diagram the proposed method

In the C-DTC method, while the stator flux vector rotation, we have to use six active voltage vector to keep stator flux vector in defined error limits, any other option. But, as described above, in SVM method, we have many options to obtain the vector what we need to control stator flux vector. The proposed fuzzy based SVM-DTC method includes a fuzzy logic controller block to produce optimum control vector. The optimum control vector angle is calculated by fuzzy logic controller with using instantaneously flux and torque errors. On this calculation, fuzzy logic rates flux and torque errors and produces necessary change in stator flux vector angle for next step. Then, calculated optimum vector angle applied to discrete space vector pulse width modulation block (DSV-PWM) and DSV-PWM generates switching signals. The Simulink block diagram of the proposed system is given in Figure 3.

The membership functions of fuzzy logic controller flux-torque inputs and angle output can be seen in Figure 4. Table 1. describes rule table of fuzzy logic controller.

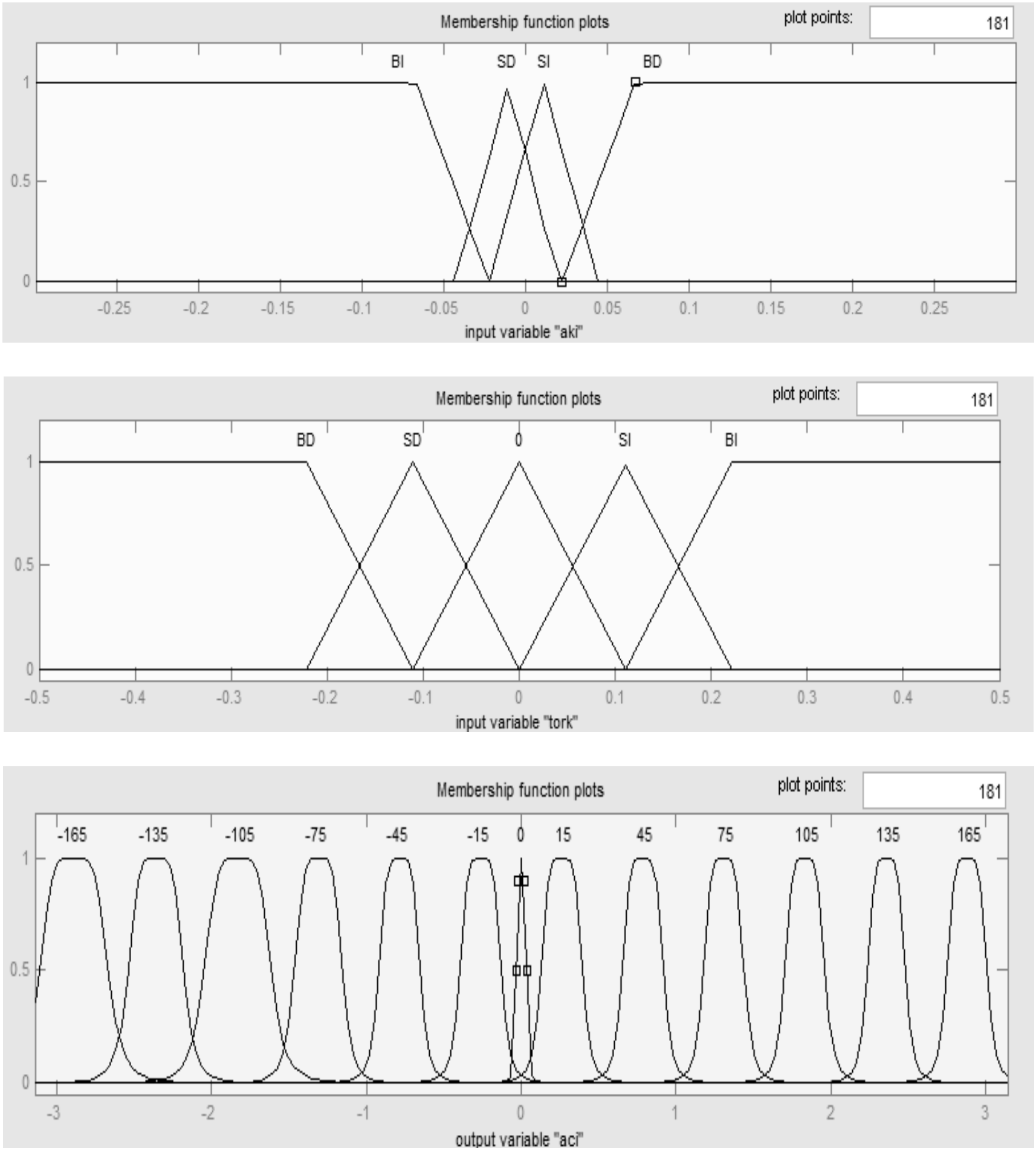

Figure 4. Flux, torque and angle membership functions

Table 1. Rule table of fuzzy logic controller

| Rules | | Flux | | | |
|---|---|---|---|---|---|
| | | BD | SD | SI | BI |
| Torque | BD | -135 | -105 | -75 | -45 |
| | SD | -165 | -135 | -45 | -15 |
| | 0 | 0 | 0 | 0 | 0 |
| | SI | 165 | 135 | 45 | 15 |
| | BI | 135 | 105 | 75 | 45 |

## 4. SIMULATION STUDIES

Numerical simulations have been carried out to evaluate the effectiveness of the proposed fuzzy logic based SVM-DTC. Its developed using Matlab/Simulink. The DTC method and the induction machine that used in the simulation works, have the parameters given in Table 2.

Table 2. Induction Machine and Simulation parameters

| IM and Simulation Parameters | |
|---|---|
| Inverter bus voltage (V) | 400V |
| Rated Power (kW) | 4 |
| Stator resistance (Ω) | 1.405 |
| Stator inductance(H) | 0.0058 |
| Pole pairs | 2 |
| Sampling time (µs) | 50 |
| Flux reference (Wb) | 0.8 |

Some tests have been carried out to compare the performances of the proposed fuzzy logic based dSV-PWM DTC (FL-dSV-PWM) with conventional DTC (C-DTC). In order to fair compare the performances with the C-DTC and the proposed FLSVM-DTC on induction motor drive, different speed and load range applied to the induction motor. The dynamic performances of the methods are performed by applying step change on load, 0Nm to 5 Nm, at 0,5. sec.

In first step of the simulation studies, the induction motor has been tested at rated speed with unloaded and loaded working conditions. The simulation results of speed and torque responses at 1500 rpm reference are shown in Figure 5. and Figure 6., respectively.

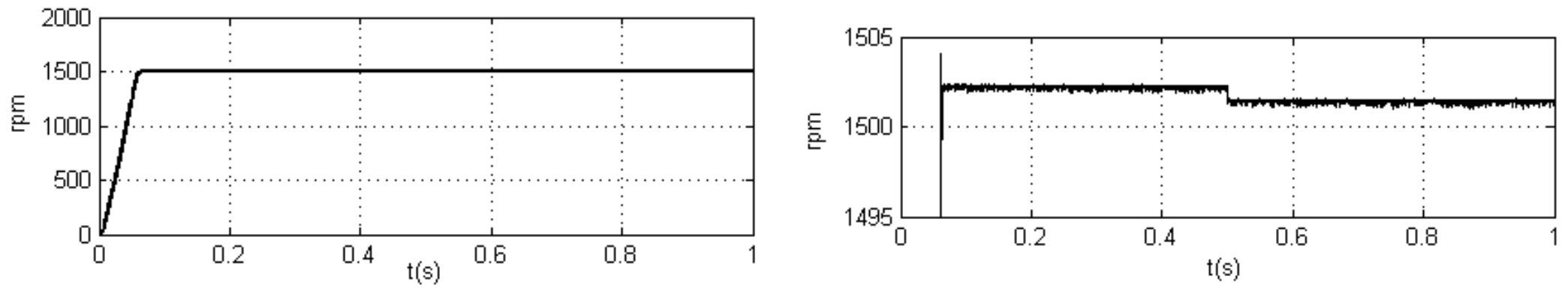

(a)

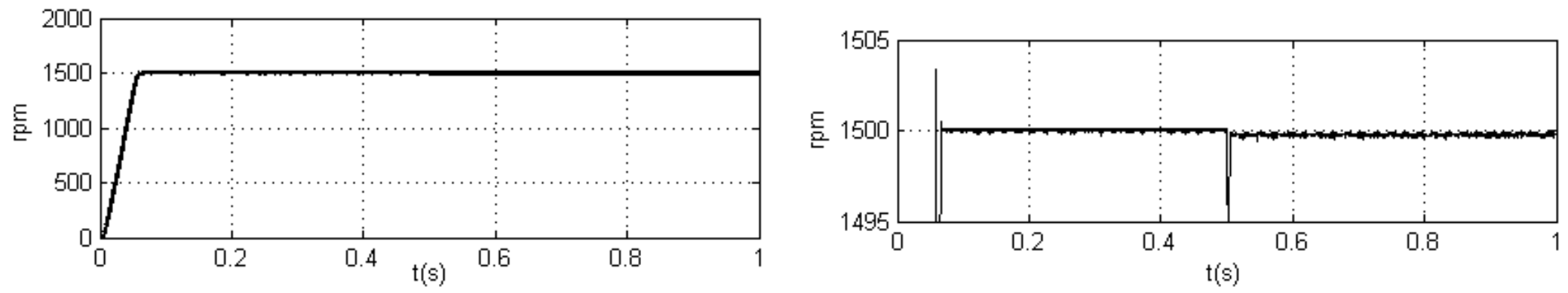

(b)

Figure 5. Speed curves of motor at 1500 rpm (with zoom view)
a) C-DTC b) FLSVM-DTC

According to the speed and torque curves which given in Figure 5. and Figure 6., the motor has reached the reference speed at about 0,075. sec. for both control method. So, it can be said that there are no difference between C-DTC and FLSVM-DTC methods speed rising times and the motor has almost same performance at transient conditions for both control method. It means, proposed control method has still fast transient conditions performance.

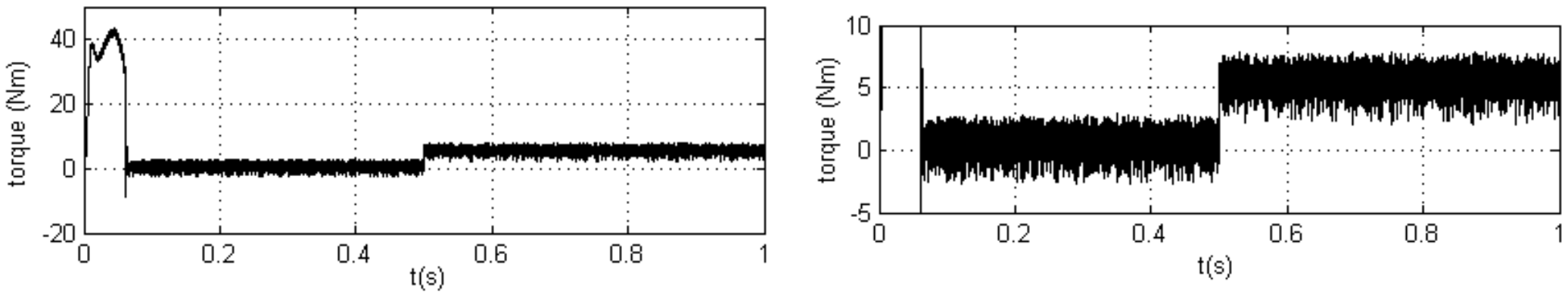

(a)

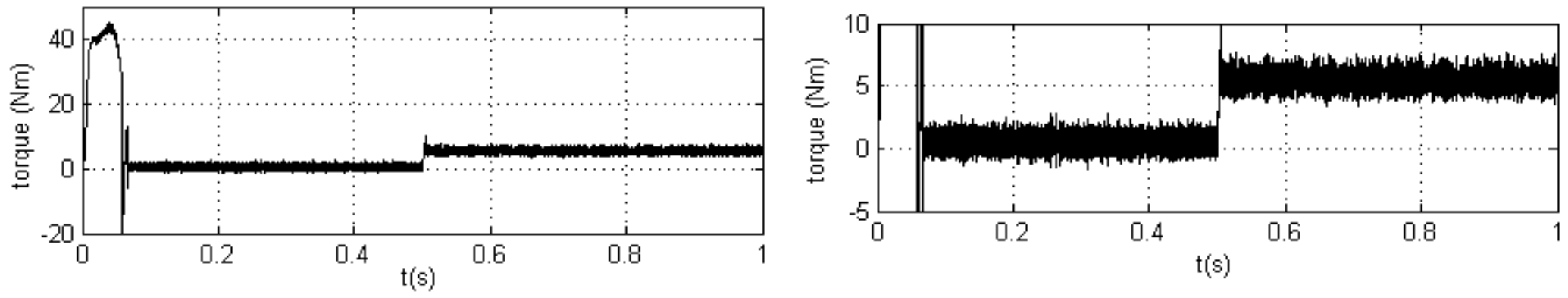

(b)

Figure 6. Torque curves of motor at 1500 rpm(with zoom view)
a) C-DTC b) FLSVM-DTC

However, the main differences have appeared at torque responses of both control methods at steady state conditions. It can be seen that, with the FLSVM-DTC instantaneous speed fluctuations and torque ripples of the motor are reduced significantly.

In the second step of simulation studies, the motor has been at low speed with two different load conditions. The aim of this tests is investigating of performances for both methods in low speed conditions. The simulation results of speed and torque responses at 250 rpm reference are shown in Figure 7. and Figure 8., respectively.

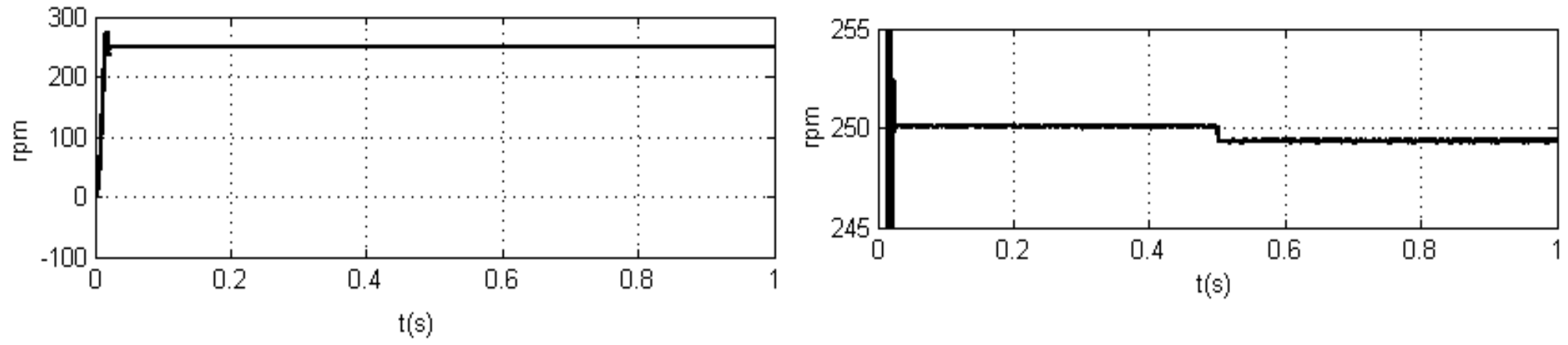

(a)

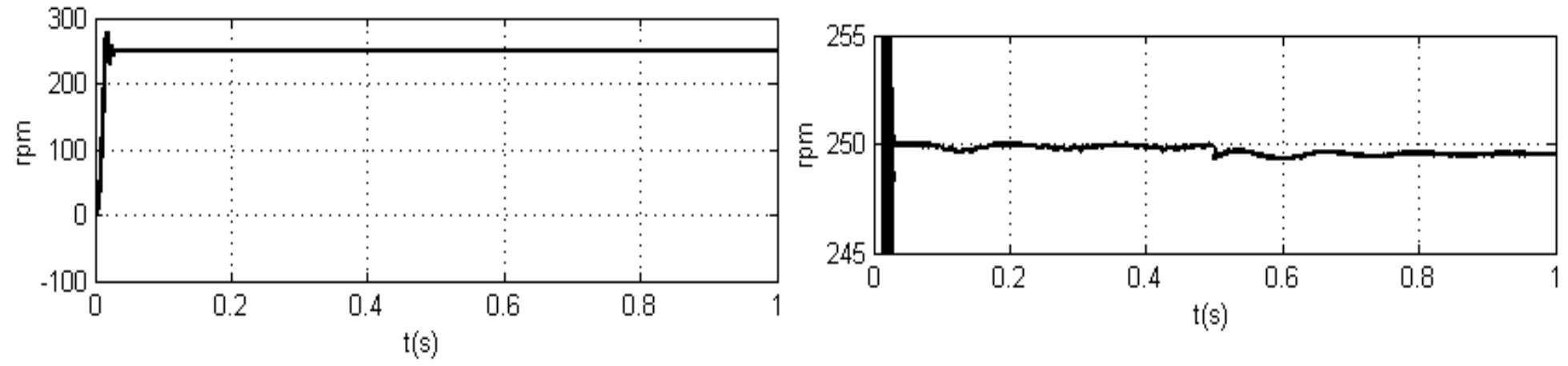

(b)

Figure 7. Speed curves of motor at 250 rpm(with zoom view)
a) C-DTC b) FLSVM-DTC

According to the speed and torque curves which given in Figure 7. and Figure 8., the motor has reached the reference speed at 0,02. sec for both control method and still no difference at transient conditions.

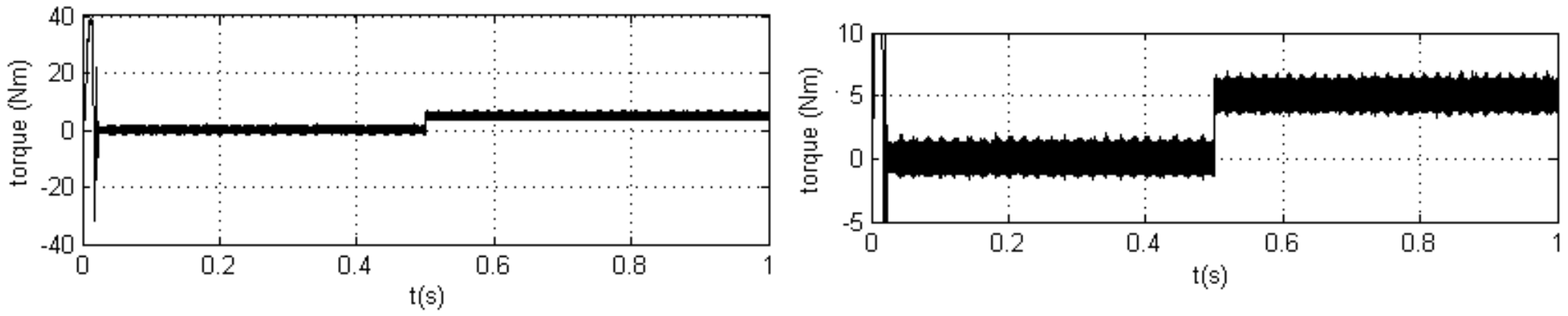

(a)

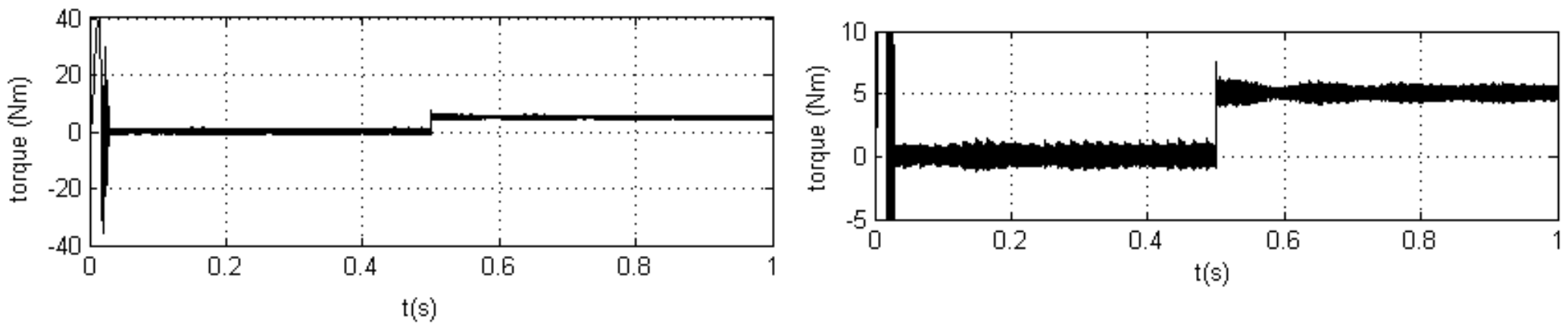

(b)

Figure 8. Torque curves of motor at 250 rpm(with zoom view)
a) C-DTC b) FLSVM-DTC

If we need to compare of steady state conditions of the motor for both control method, it must be pointed out that the FLSVM-DTC controlled motor has better performance as lesser torque ripples and speed fluctuations.

## 5. CONCLUSIONS

In this paper, a new fuzzy logic based space vector modulation technique has been proposed for the DTC controlled induction motor drivers and the simulation studies have been carried out with Matlab/Simulink to compare the proposed system behaviors at vary load and speed conditions. In the study, the induction motor has been tested for low speed (250 rpm reference) and rated speed (1500 rpm reference) working conditions to obtain fairy comparisons for both control methods. The numerical simulations prove that torque and speed responses of the motor are significantly improved with the proposed technique for both working conditions. The torque ripples of the motor are lesser about % 40, in parallel, the speed fluctuations are reduced according to conventional direct torque control technique. Moreover, the hyseteresis controllers and look-up table that used in conventional method are removed.

## Authors

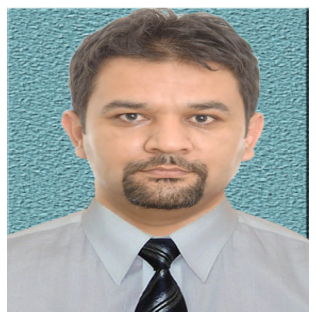

**Fatih Korkmaz** was born in Kırıkkale, Turkey, in 1977. He received the B.T., M.S., and Doctorate degrees in in electrical education, from University of Gazi,
Turkey, respectively in 2000, 2004 and 2011.
His current research field includes Electric Machines Drives and Control Systems.

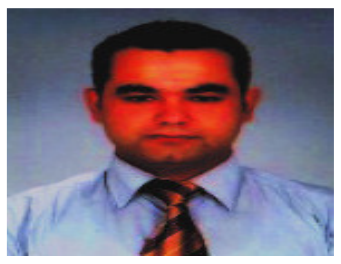

**İsmail TOPALO**GLU was born in Adana, Turkey, in 1983. He received the B.Sc , M.Sc. and Ph.D degrees in electrical education from University of Gazi in 2007,2009 and 2013, respectively.
His current research interests include Computer aided design and analysis of conventional and novel electrical and magnetic circuits of electrical machines, sensors and transducers, mechatronic systems.

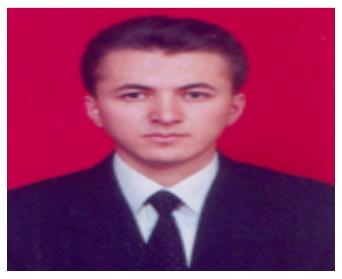

**Hayati Mamur** was born in Bolu, Turkey, in 1974. He received the B.Sc , M.Sc. and Ph.D degrees in electrical education from University of Gazi in 1996,2005 and 2013, respectively.
His research interests include automatic control, SCADA, PLC, microcontroller, DSP control applications, renewable energy, and thermoelectric modules.